\documentclass[aps,prd,twocolumn,superscriptaddress,nofootinbib,preprintnumbers]{revtex4-2}

\usepackage{amsmath,amssymb,graphicx,xcolor,hyperref}
\hypersetup{colorlinks,linkcolor=blue!60!black,citecolor=blue!60!black,urlcolor=blue!60!black}

\newcommand{\tr}{\mathrm{Tr}}
\newcommand{\erhoq}{E$\rho$OQ}
\newcommand{\ie}{\emph{i.e.}}

\begin{document}

\preprint{FERMILAB-PUB-26-0658-T}

\title{MORE Thermal Gauge Theories at Finite $\theta$ and $\mu$ \\ from Real-Time Quantum Simulation}

\author{Henry Lamm}
\affiliation{Fermi National Accelerator Laboratory, Batavia, Illinois, 60510, USA}

\date{\today}

\begin{abstract}
Imaginary-time evolution can be reconstructed from real-time quantum
simulations using exact integral transforms. We identify the
construction of Guo, Shibu, Lin, and Zhao as a continuous linear combination
of Hamiltonian simulations and show that its slow $1/t_{\rm cut}$ convergence
arises from a redundant kernel component. We extend the construction from
pure-state matrix elements to thermal traces and
correlators. One real-time dataset then reconstructs targeted inverse
temperatures, Euclidean separations, and chemical potentials through
classical post-processing, bringing finite-$(T,\theta,\mu)$ physics within
reach of real-time quantum simulation without thermal-state preparation. We
benchmark the method on one- and two-flavor lattice Schwinger
models, including circuit-level simulations with depolarizing noise.
\end{abstract}

\maketitle

\section{Introduction}
\label{sec:intro}

At nonzero topological angle $\theta$ and quark chemical potential
$\mu$, the Euclidean weight of lattice field theory (LFT) becomes complex
through the factors $e^{i\theta Q}$ and $\det M(\mu)$;
the average sign can fall as $\langle\sigma\rangle\sim e^{-V\Delta f}$,
making reweighting exponentially
costly~\cite{Troyer:2004ge,deForcrand:2010ys}. Several approaches address
the complex measure by complexifying the field
variables~\cite{Aarts:2009uq,Sexty:2013ica,Detmold:2020ncp,%
Alexandru:2020wrj} or reorganizing the path
integral~\cite{Gattringer:2016kco,Langfeld:2012ah}; analytic continuation
from imaginary $\mu$ or $\theta$ is another route, limited by
singularities such as the Roberge--Weiss
transition~\cite{deForcrand:2002hgr,DElia:2002tig,%
Borsanyi:2020fev,Dimopoulos:2021vrk,Borsanyi:2025dyp,%
Bonati:2015sqt,Roberge:1986mm}. Quantum simulations of Hamiltonians avoid
complex statistical weights because $\theta$ and $\mu$ enter through a
Hermitian Hamiltonian, but they do not directly supply the imaginary-time
evolution needed for Boltzmann factors $e^{-\beta H}$, with inverse
temperature $\beta$, or for Euclidean correlators.

\begin{figure*}[t]
  \centering
  \includegraphics[width=\textwidth]{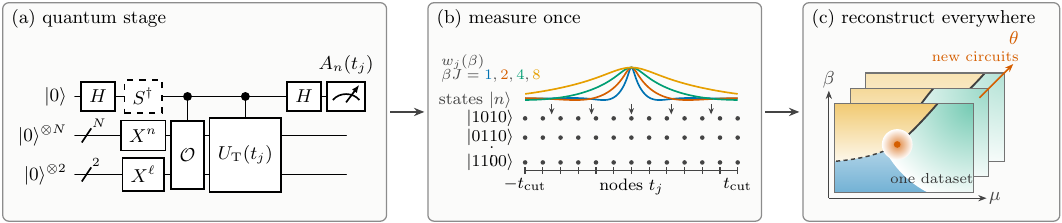}
  \caption{The \textsc{More} pipeline. (a) Quantum stage: the Hadamard
  test for one basis state $|n\rangle$ and one node $t_j$, returning
  $A_n(t_j)=\langle n|e^{-iHt_j}\mathcal{O}|n\rangle$, the diagonal case
  of the measured primitive of Eq.~\eqref{eq:trace}, from $M$ shots
  (non-unitary $\mathcal O$ is expanded in Pauli strings);
  off-diagonal elements use the same circuit on superposition
  states~\cite{Lamm:2018eroq,Saroni:2023eroq}. The optional
  $S^\dagger$ selects the imaginary part. (b) Measure once: the dataset
  is the table $A_n(t_j)$, one row per computational-basis trace state
  and one column per node of the real-time grid; the weights $w_j(\beta)$
  of Eq.~\eqref{eq:acl} are drawn for $\beta J=1,2,4,8$ ($J=g^2a/2$), so changing $\beta$ changes only the
  weights. (c) Reconstruct everywhere: $\beta$, $\tau$ and $\mu$ are
  reached by classical reweighting. Only $\theta$
  changes the Hamiltonian and requires a new dataset.}
  \label{fig:pipeline}
\end{figure*}

Guo, Shibu, Lin, and Zhao~\cite{Guo:2026ite} proposed
obtaining imaginary-time evolution from real-time simulation via the
exact operator identity
\begin{equation}
  e^{-H\tau} \;=\; \frac{i}{2\pi}\int_{-\infty}^{\infty} dt\,
  \frac{e^{ic(t+i\tau)}}{t+i\tau}\, e^{-iHt},
  \label{eq:guo}
\end{equation}
Here $\tau\ge0$ is the imaginary-time interval and $c$ is any lower
bound on the spectrum, $\mathrm{spec}(H)\subset[c,\infty)$.
Hadamard tests measure the real-time amplitudes
$\langle\psi|e^{-iHt}|\psi\rangle$, which are combined classically using
the kernel in Eq.~\eqref{eq:guo}. The imaginary-time amplitude is
thus obtained by direct quadrature of real-time data rather than by
inverse-Laplace reconstruction. Ref.~\cite{Guo:2026ite} demonstrated this
identity on quantum-mechanical toy models. Here we connect it to
standard LCU algorithms and extend it to finite-$(T,\theta,\mu)$ LFT.

We identify Eq.~\eqref{eq:guo} as a continuous linear combination of
unitaries (LCU)~\cite{Childs:2012lcu}, a form that exposes the odd kernel
component responsible for its slow convergence. Removing this component gives an exact
finite-$\ell_1$ representation, while the near-optimal kernels of An,
Childs, and Lin~\cite{An:2023acl} reduce the required real-time extent
exponentially.

Related integral and Monte Carlo representations of imaginary-time
evolution and spectral filters appear
in~\cite{Chowdhury:2016hs,Ge:2017filt,Lu:2020fine,Choi:2020rodeo,%
Somma:2019ts,Lin:2022acdf,Huo:2021mc,Zeng:2021cool,Dong:2022qetu}.
Thermal and finite-energy observables have been reconstructed from
real-time data through importance sampling of the Loschmidt
echo~\cite{Schuckert:2022jst}, constrained fits to the density of
states~\cite{Ghanem:2023rds}, kernel expansions of the density of
states~\cite{Wang:2022hcd}, and microcanonical
filtering~\cite{Hemery:2023hh}. The contemporaneous
Ref.~\cite{Tang:2026msqite} instead optimizes the energy-shift freedom in
Eq.~\eqref{eq:guo}.

We extend the method
to partition functions, thermal traces, and Euclidean correlators within
\erhoq{}, the framework for evolving density matrices on
qubits~\cite{Lamm:2018eroq,Saroni:2023eroq}. The kernel supplies the
Boltzmann weights from the same real-time circuits, replacing the classical
imaginary-time input on which earlier implementations
relied~\cite{Harmalkar:2020mpq,Gustafson:2020vqz} and where sign problems
re-enter at $\theta,\mu\neq0$~\cite{Gustafson:2026lcu}.

The resulting method, which we call \textsc{More} for \emph{measure
once, reconstruct everywhere}, is summarized in Fig.~\ref{fig:pipeline}:
real-time amplitudes are measured on a fixed grid $\{t_j\}$ and combined
classically with weights $w_j$ that carry the entire $(\beta,\tau,\mu)$
dependence.

Thermal LFT observables have so far been largely obtained from quantum
algorithms that \emph{prepare} the thermal state in classical emulation,
through variational
free energies~\cite{Wu:2018nrn,Xie:2022jgj,Tomiya:2022bvqe,Fromm:2023npm,Cheng:2024pdu}, thermal pure quantum
states~\cite{Davoudi:2022pcz}, minimally entangled typical thermal
states~\cite{Chen:2024oao,Maeno:2026ooq}, quantum imaginary-time
evolution~\cite{Motta:2019yya,Czajka:2021yll,Pedersen:2023tpq,Ikeda:2024rzv}, or quantum
Metropolis and Gibbs~\cite{Temme:2009wa,QuBiPF:2020iiz,Ballini:2023qms,Chen:2023cuc} sampling, with variational thermal
states of one-dimensional $SU(2)$ and $SU(3)$ prepared on trapped-ion
hardware~\cite{Than:2024zaj}; \erhoq{} itself has produced
thermal results from classical imaginary-time
input~\cite{Lamm:2018eroq,Saroni:2023eroq}, including quenches of the PXP
model on superconducting hardware~\cite{Desaules:2023yhw}.

Adiabatic
preparation of the $\theta$-vacuum was studied numerically
in~\cite{Chakraborty:2020theta,Honda:2021nvv,Kaikov:2024acw}; the critical
endpoint~\cite{Thompson:2021eze} and first-order
transition~\cite{Angelides:2023fot} at $\theta=\pi$ were then studied on
superconducting hardware, and a tunable $\theta$-angle was
proposed~\cite{Halimeh:2022pkw} and realized~\cite{Zhang:2023theta} in a
cold-atom simulator, with real-time scattering at tunable $\theta$ run on
superconducting hardware~\cite{Schuhmacher:2025ehh}. At finite
density, ground states and the phase structure have been studied
variationally~\cite{Yamamoto:2021vxp,Schuster:2023klj}. Few
works treat both $\theta$ and $\mu$: the phase diagram located by
entanglement in~\cite{Ikeda:2023crit}, and the three-dimensional SU(3)
demonstration of~\cite{Hidalgo:2026lqcd}, where $\theta$ enters through
real-time evolution and $\mu$ only as an energy difference between
baryon-number sectors.

Unlike methods that prepare separate thermal states,
\textsc{More} reuses one real-time dataset across all targeted values of
$\beta$, $\tau$, and $\mu$.
We demonstrate the pipeline on the one-flavor
lattice Schwinger model at finite $(T,\theta)$ and the two-flavor model
at finite isospin density, benchmarked against exact diagonalization.

\section{$\tau$-kernels and costs}
\label{sec:lcu}

On an eigenstate of energy $E$, Eq.~\eqref{eq:guo} reduces to the Fourier integral
\begin{equation}
  \frac{i}{2\pi}\int dt\, \frac{e^{-ixt}}{t+i\tau}
  = \Theta(x)\,e^{-x\tau}, \qquad x = E - c,
  \label{eq:pair}
\end{equation}
\ie{} the one-sided filter $\Theta(E-c)e^{-(E-c)\tau}$, times
$e^{-c\tau}$. Decompose the kernel into even and odd parts,
\begin{equation}
  \frac{i}{2\pi}\frac{1}{t+i\tau}
  \;=\; \underbrace{\frac{1}{2}\,\frac{\tau/\pi}{t^2+\tau^2}}_{\text{even part}}
  \;+\; \underbrace{\frac{i}{2}\,\frac{t/\pi}{t^2+\tau^2}}_{\text{odd part}} .
  \label{eq:split}
\end{equation}
The Fourier transforms of the even and odd parts are, respectively,
$\tfrac12 e^{-|x|\tau}$ and
$\tfrac12\operatorname{sgn}(x)e^{-|x|\tau}$, which add to the
one-sided filter $\Theta(x)e^{-x\tau}$. Throughout, $t$ denotes physical
real time and $k=t/\tau$ its measure in units of the imaginary time
being reconstructed; kernel shapes and convergence rates are
functions of $k$, while circuit depth is set by $t$. In the variable $k$,
twice the even part is the Cauchy kernel
$\frac{1}{\pi(1+k^2)}$ of the linear-combination-of-Hamiltonian-simulation
(LCHS) construction of Ref.~\cite{An:2023lchs}; specialized to a
Hermitian, time-independent generator, that construction is the
imaginary-time representation of Refs.~\cite{Zeng:2021cool,Huo:2021mc}. On
the physical spectrum $x\geq0$ the even part alone reproduces
$e^{-x\tau}$, so the odd part is redundant. Removing it gives the exact representation
\begin{equation}
  e^{-H\tau} \;=\; e^{-c\tau} \int_{-\infty}^{\infty} dt\;
  \frac{\tau/\pi}{t^{2}+\tau^{2}}\; e^{-i(H-c)t}.
  \label{eq:poisson}
\end{equation}
The near-optimal kernel of An, Childs, and
Lin~\cite{An:2023acl} gives
\begin{equation}
  e^{-(H-c)\tau} = \!\int\! dk\, \frac{f(k)}{1-ik}\, e^{-i(H-c)k\tau},
  \quad f(k)=\frac{e^{-(1+ik)^{\alpha}}}{C_{\alpha}},
  \label{eq:acl}
\end{equation}
where the integration variable is $k$ rather than $t$, $0<\alpha<1$,
and $C_{\alpha}=2\pi e^{-2^{\alpha}}$ is fixed by the normalization
$\int dk\, f(k)/(1-ik)=1$.

\begin{figure}[t]
  \centering
  \includegraphics[width=\linewidth]{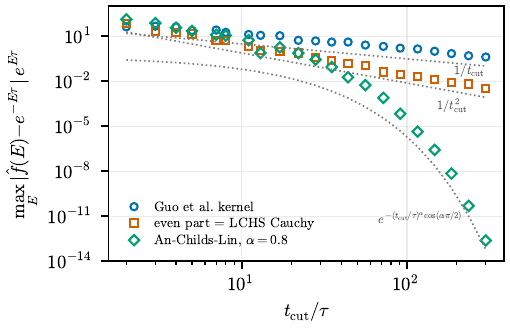}
  \caption{Worst-case relative error of the reconstructed filter
  $e^{-E\tau}$ over the spectrum versus $t_{\rm cut}/\tau$, for the
  kernels of Eq.~\eqref{eq:guo}~\cite{Guo:2026ite}, Eq.~\eqref{eq:poisson},
  and Eq.~\eqref{eq:acl}~\cite{An:2023acl}. Dotted lines and their
  adjacent labels show the predicted convergence laws.}
  \label{fig:kernels}
\end{figure}

Truncated to $|t|\le t_{\rm cut}$ and discretized on quadrature nodes
$t_j$ with weights $w_j$, all three kernels realize the approximate
filter
\begin{equation}
  \hat f(E)=\sum_j w_j e^{-i(E-c)t_j}.
  \label{eq:filter}
\end{equation}
We quantify kernel error as the deviation of $\hat f$ from the target filter. For
the original kernel, the redundant odd component has a $1/|t|$ tail that
causes both a logarithmically divergent $\ell_1$ norm and the
$1/t_{\rm cut}$ truncation error derived in~\cite{Guo:2026ite}. The even
kernel instead has finite $\ell_1=e^{-c\tau}$ and truncation error
$O(\tau/[(E-c)\,t_{\rm cut}^{2}])$. Eq.~\eqref{eq:acl} converges near-exponentially.
Let $\epsilon_{\rm trunc}$ be the target operator-norm error from truncating the
kernel integral, and define the required dimensionless cutoff by
\begin{align}
  K(\epsilon_{\rm trunc})&\equiv
  \inf\left\{K>0:
  \int_{|k|>K}\!dk\,
  \left|\frac{f(k)}{1-ik}\right|\le\epsilon_{\rm trunc}\right\} \nonumber\\
  &=O\!\left(\log^{1/\alpha}\frac{1}{\epsilon_{\rm trunc}}\right),
  \qquad
  t_{\max}=\tau K(\epsilon_{\rm trunc}).
  \label{eq:kernel-reach}
\end{align}
Here the scaling follows from
$|f(k)|\le C_\alpha^{-1}e^{-|k|^{\alpha}\cos(\alpha\pi/2)}$ and improves
exponentially on the $O(1/\epsilon_{\rm trunc})$ time required by the original
kernel. A Phragm\'en--Lindel\"of argument shows the choice is
near-optimal within that family~\cite{An:2023acl}.
Fig.~\ref{fig:kernels} compares the worst-case filter error as a
function of $t_{\rm cut}/\tau$ for the three kernels.

Each kernel requires a lower bound $c\leq E_{\min}$. A conservative bound uses
$E_{\min}\geq-\lVert H\rVert$. A loose bound preserves
exactness but increases the overlap penalty $\lVert w\rVert_1/Z\propto
e^{\beta(F-c)}$, with $F$ the free energy, and thus the shot cost. We take
$c=E_{\min}-\delta$ with $\delta=0.5J$ from ED; in practice $c$ must be within
$O(1/\beta)$ of $E_{\min}$.

\label{sec:cost}

\begin{figure}[t]
  \centering
  \includegraphics[width=\linewidth]{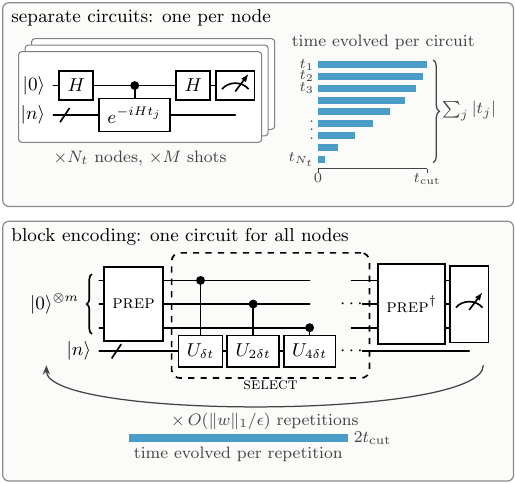}
  \caption{Two implementations of $\sum_j w_j e^{-iHt_j}$. The
  separate-circuit approach (top) uses $N_t$ Hadamard-test circuits, $M$
  shots, and $R$ trace states, for $N_tRM$ executions, with variance at most
  $\|w\|_1^2/M$. The block-encoding approach (bottom) stores the weights in
  an $m=\lceil\log_2N_t\rceil$-qubit address register and applies the
  controlled evolutions, with each query spanning at most $2t_{\rm cut}$.}
  \label{fig:lcu}
\end{figure}

Discretized on quadrature nodes, each of Eqs.~\eqref{eq:guo},
\eqref{eq:poisson}, and~\eqref{eq:acl} is a standard LCU
$\sum_j w_j e^{-iHt_j}$. In the separate-circuit implementation used
in~\cite{Guo:2026ite} and here, each amplitude is measured with a
Hadamard test. Alternatively, the
weighted sum can be implemented as a block encoding using
\textsc{prepare}/\textsc{select} and amplitude
amplification~\cite{Childs:2012lcu,An:2023acl} (Fig.~\ref{fig:lcu}). In
the single-ancilla LCU
framework~\cite{Chakraborty:2023lcu,Wang:2023rand}, the separate-circuit
estimator has variance $\propto\|w\|_1^2/M$ with $M$ total shots.
Thermal expectation values then inherit the squared penalty
$e^{2\beta(F-c)}$ in their variance. For hardware resources, one useful metric is
the usage-weighted total evolved time $\sum_j |t_j|$; the $\alpha<1$
kernels compress this cost near-optimally.

Quantum simulations incur Trotter error $\epsilon_{\rm T}$ and hardware error
$\epsilon_{\rm noise}$. We model the latter as depolarizing noise parameterized by circuit volume,
$A(t)\to e^{-\gamma n_{\rm CX}(t)}A(t)$, where $n_{\rm CX}(t)$ is the
transpiled CX count and $\gamma$ is an effective per-gate damping rate.
For evolution to $t$, we use
$n_{\rm T}(t)=\max(1,\lfloor r_{\rm T}|t|\rfloor)$ second-order Trotter
steps, where $r_{\rm T}$ is the number of steps per unit simulation time.
Because $n_{\rm T}(t)$ is discrete, $n_{\rm CX}(t)$ is a staircase in
$|t|$. For analytic interpretation, we approximate this by
$e^{-\lambda|t|}$. It modifies Eq.~\eqref{eq:filter} by convolution with a Lorentzian of width $\lambda$:
\begin{equation}
  \hat f_\lambda(E) = \int dE'\, \frac{\lambda/\pi}{(E-E')^2+\lambda^2}\,
  \hat f(E').
  \label{eq:noise}
\end{equation}
This relation motivates a time-domain correction:
the measured amplitudes are reweighted by the inverse fitted envelope
before quadrature. Since $t_{\max}\propto\beta$, damping
limits $\beta$ to $O(1/\lambda)$.

\section{Thermal traces and Euclidean correlators in \erhoq{}}
\label{sec:thermal}

Phase diagrams require traces, with periodic boundary conditions in
Euclidean time and antiperiodic ones for fermions, not pure-state matrix
elements. The \erhoq{} framework~\cite{Lamm:2018eroq,Saroni:2023eroq}
evaluates
\begin{equation}
  \tr[e^{-\beta H}\mathcal{O}]
  = \sum_{mn} \langle m|e^{-\beta H}|n\rangle\langle n|\mathcal{O}|m\rangle
\end{equation}
over an ensemble of superpositions of computational-basis states;
diagonal measurements on those superpositions recover the off-diagonal
matrix elements~\cite{Lamm:2018eroq,Saroni:2023eroq}. Previously the density matrix
$\langle m|e^{-\beta H}|n\rangle$ came from classical
imaginary-time input. Here instead every Boltzmann-weighted matrix element is
    reconstructed from the measured data,
\begin{equation}
  \langle n| e^{-\beta H} \mathcal{O} |m\rangle
  = \int dt\; K_\beta(t)\,
    \langle n| e^{-iHt}\, \mathcal{O} |m\rangle ,
  \label{eq:trace}
\end{equation}
with $K_\beta$ any kernel of Sec.~\ref{sec:lcu}. Thermal expectation
values are the normalized traces,
\begin{equation}
  \langle\mathcal O\rangle_\beta
  =\frac{\tr[e^{-\beta H}\mathcal O]}{\tr[e^{-\beta H}]},
  \label{eq:expval}
\end{equation}
with $Z=\tr[e^{-\beta H}]$ the partition function.
Euclidean correlators follow from two insertions,
\begin{align}
  G_{\mathcal O_1\mathcal O_2}(\tau) &\equiv
             \frac{1}{Z}\tr\!\big[e^{-(\beta-\tau)H}\mathcal{O}_1
             e^{-\tau H}\mathcal{O}_2\big] \nonumber\\
  &= \frac{1}{Z}\sum_n \int\! dt_1\, dt_2\;
    K_{\beta-\tau}(t_1)\, K_{\tau}(t_2) \nonumber\\
  &\qquad\times\;
    \langle n| e^{-iHt_1}\mathcal{O}_1 e^{-iHt_2}\mathcal{O}_2|n\rangle,
  \label{eq:kms}
\end{align}
If a symmetry $U$ commutes with both $H$ and $\mathcal O$, basis
states in the same $U$-orbit give identical diagonal matrix elements. The
trace can therefore be evaluated using one representative from each orbit,
weighted by the orbit size, rather than every basis state independently,
without changing the quantum primitive~\cite{Saroni:2023eroq}.

Thermal correlators obey the Kubo--Martin--Schwinger (KMS)
relation~\cite{Kubo:1957mj,Martin:1959jp},
\begin{equation}
  G_{\mathcal{O}_1\mathcal{O}_2}(\beta-\tau)
  = G_{\mathcal{O}_2\mathcal{O}_1}(\tau),
  \label{eq:kmsrel}
\end{equation}
By cyclicity of the trace, Eq.~\eqref{eq:kmsrel} is an identity for the
complete trace ensemble, whatever the truncation and quadrature errors, so
for sampled ensembles its residual measures ensemble incompleteness and
shot noise.

\section{$\theta$-terms and chemical potentials}
\label{sec:thetamu}

Both couplings become Hermitian Hamiltonian terms, but at different
cost. $\theta$ modifies $H$, and
$[H(\theta),H(\theta')]\neq0$ for $\theta\neq\theta'$, so
every $\theta$ requires a separate dataset. By contrast, $\mu$ couples
to a conserved charge and thus commutes out of the
Boltzmann factor, so the whole $\mu$ axis can in favorable cases be
recovered from a single $\mu=0$ dataset by classical reweighting. We
take the two in turn, specializing to $1{+}1$d $U(1)$ with staggered
fermions.

In the one-flavor Schwinger model, an anomalous chiral rotation shifts
$\theta$ into the mass term,
\begin{equation}
  m\bar\psi\psi \;\longrightarrow\;
  m\bar\psi e^{i\gamma_5\theta}\psi .
\end{equation}
For staggered fermions, the additive
mass counterterm $m_{\rm lat}=m-g^2a/8$ is
required for the lattice to realize the continuum symmetry
$(m,\theta)\to(-m,\theta+\pi)$~\cite{Dempsey:2022nys}. This
leads to the Hamiltonian of the lattice Schwinger model~\cite{Chakraborty:2020theta}:
\begin{align}
  H(\theta) &= \sum_{n}\Big(\tfrac{1}{2a}
      + (-1)^n \tfrac{m}{2}\sin\theta\Big)
      \big(\chi^\dagger_n U_n \chi_{n+1} + \text{h.c.}\big) \nonumber\\
  &\quad + \sum_n \big(m\cos\theta - \tfrac{g^2a}{8}\big)(-1)^n
      \chi^\dagger_n\chi_n
    + \frac{g^2 a}{2}\sum_{n} L_n^{2}.
  \label{eq:schwinger}
\end{align}
With periodic boundaries, Gauss's law $L_n-L_{n-1}=Q_n$, where
$Q_n=\chi_n^\dagger\chi_n-\tfrac12[1-(-1)^n]$, leaves one
integer-valued holonomy $\ell$:
\begin{equation}
  L_n=\ell+\sum_{k\le n}Q_k,\qquad \sum_n Q_n=0.
  \label{eq:pbcgauss}
\end{equation}
We truncate it to $|\ell|\le\ell_{\max}$, giving a physical-space
dimension $(2\ell_{\max}+1)\binom{N}{N/2}$. The wrap-around hopping shifts $\ell$ and
retains its Jordan--Wigner string. Because of how $\theta$ enters
Eq.~\eqref{eq:schwinger},
$H(\theta+2\pi)=H(\theta)$ at any truncation. We characterize
the finite-temperature $\theta$ dependence using the topological charge,
\begin{align*}
  \langle Q\rangle
  =\Big\langle\sum_n(-1)^n\Big[&\tfrac{m}{2}\cos\theta\,
    \big(\chi^\dagger_n U_n\chi_{n+1}+\text{h.c.}\big)\\
    &-m\sin\theta\,\chi^\dagger_n\chi_n\Big]\Big\rangle ,
\end{align*}
and the chiral condensate ($\bar\psi e^{i\gamma_5\theta}\psi$ in the unrotated fields),
\begin{equation}
  \langle\bar\psi\psi\rangle
  =\frac{1}{N}\sum_n(-1)^n
    \langle\chi_n^\dagger\chi_n\rangle.
  \label{eq:condensate}
\end{equation}

\label{sec:ladder}

In contrast to $\theta$-terms, $\mu$ couples to a conserved charge
operator $\hat N$, such as
baryon number, electric charge, strangeness, or the third component of
isospin. In the two-flavor extension of Eq.~\eqref{eq:schwinger}, each
site carries two species, and the isospin charge
$I_3=\tfrac12\sum_n(\hat n_{n,0}-\hat n_{n,1})$ enters through
$H-\mu_I I_3$. Removing $\mu$ from the Hamiltonian simulation relies on
three properties that hold in standard lattice formulations.

First, charge conservation implies $[H,\hat N]=0$, so
\begin{equation*}
  e^{-\beta(H-\mu\hat N)}=e^{-\beta H}e^{\beta\mu\hat N}.
\end{equation*}
The grand-canonical trace thus decomposes into canonical charge sectors
weighted by the fugacity factor $z^{N_q}=e^{\beta\mu N_q}$, where
$z=e^{\beta\mu}$. This is the Hamiltonian counterpart of the classical
canonical formalism~\cite{Hasenfratz:1991ax,Alexandru:2005ix}.

Second, because the charge is a sum of commuting local terms,
$\hat N = \sum_x \hat n_x$ with $[\hat n_x,\hat n_y]=0$,
$e^{i\phi\hat N}=\prod_x e^{i\phi\hat n_x}$ factorizes into a single
layer of one-site rotations. The canonical projector
\begin{equation}
  \hat P_N = \int_0^{2\pi}\frac{d\phi}{2\pi}\,
             e^{i\phi(\hat N - N)}
  \label{eq:proj}
\end{equation}
is then affordable: rather than building $\hat P_N$ as a circuit, we draw $\phi$
uniformly at each shot and apply the single layer $e^{i\phi\hat N}$
within the circuit, so the projection costs one layer of depth. The continuation
$\phi\to-i\beta\mu$ then reaches real $\mu$. This is an exact quantum
counterpart of the imaginary-$\mu$
method~\cite{deForcrand:2002hgr}, but it does not rely on fitting an
analytic continuation. In the conventional approach, the
Roberge--Weiss transition bounds the continuation domain in the
thermodynamic limit~\cite{Roberge:1986mm}, and the continuation is fitted
within that domain~\cite{deForcrand:2002hgr,DElia:2002tig}. Here, at fixed
finite volume the charge is bounded, so $Z$ is a polynomial in the
fugacity with no singularities, and $\phi$ is integrated as a projector
rather than used as fit data. The practical limitation is
thus variance, not a radius of convergence.

Finally, for occupation-number trace states,
$\hat N|n\rangle=N_n|n\rangle$, $\mu$ drops out of the quantum matrix
element:
\begin{equation}
  \tr\big[e^{-\beta(H-\mu \hat N)}\mathcal{O}\big]
  = \sum_n e^{\beta\mu N_n}\,
    \langle n|\, e^{-\beta H}\mathcal{O} |n\rangle ,
  \label{eq:fugacity}
\end{equation}
Thus $\mu$ enters through the positive classical factor
$e^{\beta\mu N_n}$ multiplying each reconstructed matrix element. The
spectral bound $c$ is therefore $\mu$-independent.

Positivity removes the sign problem but not the overlap problem. As the
fugacity weights concentrate in high-charge sectors, the effective
sample size $N_{\rm eff}=(\sum_n w_n)^2/\sum_n w_n^2$ falls toward the
multiplicity of the dominant sector. A solution is that because
$e^{\beta\mu N_n}$ factorizes over sites, it can instead define the
sampling distribution: drawing each site occupied with probability
$e^{\beta\mu}/(1+e^{\beta\mu})$ reproduces the weight $e^{\beta\mu N_n}$
in the unconstrained Fock space; with Gauss's law the draws must be
conditioned on the physical sector. This mirrors Euclidean canonical
sampling~\cite{Hasenfratz:1991ax,Alexandru:2005ix}.

Fugacity reweighting changes which sector errors control the total
accuracy. Let $Z_q=\tr_{\mathcal H_q}e^{-\beta H}$, where
$\mathcal H_q$ has charge $N_q$, and let
$\widehat Z_q=Z_q(1+\epsilon_q)$, where $\epsilon_q$ is the relative
error in the $q$-th sector partition function. Then
\begin{equation}
  \frac{\widehat Z(\mu)-Z(\mu)}{Z(\mu)}=\sum_q p_q(\mu)\epsilon_q,
  \quad p_q(\mu)=\frac{e^{\beta\mu N_q}Z_q}{Z(\mu)},
  \label{eq:sector-error}
\end{equation}
with $\sum_q p_q(\mu)=1$.
Increasing $\mu$ can make sectors with large
$\epsilon_q$ important. Because the spectral bound enters
only through the classical reconstruction weights, each sector may
instead use $c_q=E_{\min}^{(q)}-\delta$, without requiring new quantum
data. Sector-resolved shifts thereby improve accuracy in the sectors
where $p_q(\mu)$ is appreciable.

The Silver-Blaze phenomenon~\cite{Cohen:2003kd}, the exact
$\mu$-independence of observables below the gap at $T\to0$, becomes in
Eq.~\eqref{eq:fugacity} a statement about classical reweighting of one
dataset, and hence a test of the reconstruction.

\begin{figure*}[t]
  \centering
  \includegraphics[width=\textwidth]{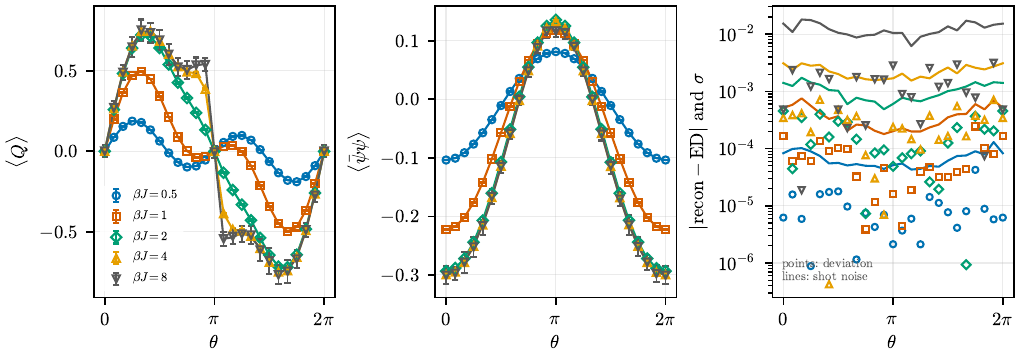}
  \caption{One-flavor Schwinger model with periodic boundaries, $N=10$,
  holonomy truncation $\ell_{\max}=1$, at five temperatures: topological
  charge $\langle Q\rangle$ (left) and chiral
  condensate $\langle\bar\psi\psi\rangle$ (center), reconstructed from real-time data (open markers,
  one shape per temperature; means over 30 resamplings, bars the spread
  of one) against ED (lines). The
  right panel shows, for $\langle\bar\psi\psi\rangle$,
  $|\text{recon}-\text{ED}|$ as points and the shot-noise spread
  $\sigma$ as lines.}
  \label{fig:theta}
\end{figure*}

\section{Demonstrations}
\label{sec:results}

We demonstrate the complete pipeline on the one-flavor
model of Eq.~\eqref{eq:schwinger} and its two-flavor extension.
We use units $J=g^2a/2$, with $x=1/(ga)^2=1$ and $m/g=0.3$.
All reconstructions use the ACL kernel with $\alpha=0.8$ on
Gauss--Legendre nodes.
Except in the correlator and noise studies, exact
statevector amplitudes provide the
noiseless means, and finite-shot fluctuations are added by resampling each
circuit at $10^6$ shots; error bars are the standard deviation over $30$
such resamplings. We benchmark against exact diagonalization (ED).

We decompose the reconstruction error into the budget
\begin{equation}
  \epsilon_{\rm tot}\lesssim
  \epsilon_{\rm trunc}+\epsilon_{\rm disc}+\epsilon_{\rm T}
  +\epsilon_{\rm noise}+\epsilon_{\rm stat}+\epsilon_{\rm syn}.
  \label{eq:error-budget}
\end{equation}
Increasing $t_{\rm cut}$ suppresses $\epsilon_{\rm trunc}$, increasing
$N_t$ suppresses $\epsilon_{\rm disc}$, and increasing $r_{\rm T}$
suppresses $\epsilon_{\rm T}$ but increases circuit depth and hence
$\epsilon_{\rm noise}$. $\epsilon_{\rm stat}$ decreases with the
shot and trace-state budgets, while $\epsilon_{\rm syn}$ is fixed by the
chosen rotation-synthesis tolerance.

Two conditions set the number of quadrature nodes $N_t$. First, the
measured amplitudes oscillate at frequencies up to the spectral width
$\Delta E=E_{\max}-E_{\min}$ of $H$, which includes high-energy
configurations with large electric flux, and resolving them over
$|t|\le t_{\rm cut}$ requires
\begin{equation}
  N_t\gtrsim\frac{3.3\,\Delta E\,t_{\rm cut}}{\pi},
  \label{eq:frequency-resolution}
\end{equation}
about $3.3$ nodes per period of $e^{-i\Delta E t}$ on average:
Gauss--Legendre nodes are sparsest at the center of the grid, where their
spacing is $\pi t_{\rm cut}/N_t$, so the Nyquist condition there requires
$\pi$ per period on average, and the remaining $5\%$ is margin. Here $t_{\rm cut}$ is set by
the largest targeted $\beta$. Second, the kernel
must be resolved, and here $\epsilon_{\rm trunc}$ and $\epsilon_{\rm disc}$
compete at fixed $N_t$: increasing $t_{\rm cut}$ suppresses
$\epsilon_{\rm trunc}$, but spreads the same nodes over a wider real-time
interval and can therefore increase $\epsilon_{\rm disc}$. The kernel is
truncated unless
$t_{\rm cut}\gtrsim \max(\tau,\beta-\tau)\,K(\epsilon_{\rm trunc})$, with
$K(\epsilon_{\rm trunc})$ defined in Eq.~\eqref{eq:kernel-reach}. But
$K_\tau$ carries weight only over $|t|\lesssim\tau K(\epsilon_{\rm trunc})$,
so on a grid spanning $\pm t_{\rm cut}$ only a fraction
$\tau K/t_{\rm cut}$ of the nodes do any work. With $\mathcal C$ nodes needed
per unit $\tau$ of that support, resolving the shorter Euclidean segment of
a correlator requires
\begin{equation}
  N_t \;\gtrsim\; \mathcal C\,\frac{t_{\rm cut}}{\min(\tau,\,\beta-\tau)}
  \;\sim\; \mathcal C\,K(\epsilon_{\rm trunc})\,
  \frac{\max(\tau,\beta-\tau)}{\min(\tau,\beta-\tau)} .
  \label{eq:resolution}
\end{equation}
The cost is set by the ratio of the two segments. A thermal trace applies one kernel, but a grid
shared by $\beta_{\min}\le\beta\le\beta_{\max}$ has the same imbalance:
Eq.~\eqref{eq:resolution} becomes
$N_t\gtrsim\mathcal C\,K(\epsilon_{\rm trunc})\beta_{\max}/\beta_{\min}$.
The larger of the two lower bounds in Eqs.~\eqref{eq:frequency-resolution}
and~\eqref{eq:resolution} determines $N_t$: for the trace of
Fig.~\ref{fig:theta} it is the first ($N_t\approx3.7\times10^4$ at
$t_{\rm cut}=480$); for the correlator below, the second.

Fig.~\ref{fig:theta} shows the topological charge $\langle Q\rangle$
and $\langle\bar\psi\psi\rangle$ for $N=10$ at $\beta J=0.5,1,2,4,8$. Across all
reconstructed points, the deviation from ED never exceeds $0.49$ times
the statistical spread, showing that shot noise rather than kernel error
limits the reconstruction. The expected symmetries emerge: $\langle Q\rangle$ is $CP$-odd and vanishes at
$\theta=0,\pi,2\pi$ to $2\times10^{-3}$ at $\ell_{\max}=1$, while $\langle\bar\psi\psi\rangle$ is $2\pi$-periodic at any $\ell_{\max}$ and $CP$-even to $5\times10^{-5}$
at $\ell_{\max}=1$, comparable to the smallest error bar, and exact for
$\ell_{\max}\ge2$.

As the temperature falls, $\langle Q\rangle$ steepens near
$\theta=\pi$, while $\langle\bar\psi\psi\rangle$ changes little between $\beta J=4$
and $8$. At zero temperature, the massive Schwinger model has a
first-order line at $\theta=\pi$ terminating at an Ising
endpoint~\cite{Byrnes:2002gj,Shimizu:2014uva}, located in the continuum at
$(m/g)_c=0.333556(5)$~\cite{Fujii:2024tjb}. Our choice $m/g=0.3$ lies below this endpoint, where $\theta=\pi$ has no
transition even at $T=0$~\cite{Ohata:2023sqc}.
Since the lattice is small and coarse, our results are not a continuum
prediction.

Related observables have been computed with
GTRG~\cite{Kanno:2024grg}, matrix product
operators~\cite{Banuls:2015sta,Banuls:2016lkq,Buyens:2016ecr}, and thermal-pure-state
QITE~\cite{Pedersen:2023tpq}.

Across $\beta J=0.5$--$8$, $\epsilon_{\rm stat}$ tracks
$\lVert w\rVert_1/Z\propto e^{\beta(F-c)}$, with $F$ the free energy, to within a
factor of three while growing $150$--$200$-fold. The bound
$\lVert w\rVert_1\sqrt D/(\sqrt M Z)$, for trace dimension $D$, is one to two orders of magnitude looser because $\epsilon_{\rm stat}$
adds in quadrature across the nodes and partially cancels between
the numerator and denominator of Eq.~\eqref{eq:expval}.

\begin{figure}[t]
  \centering
  \includegraphics[width=\linewidth]{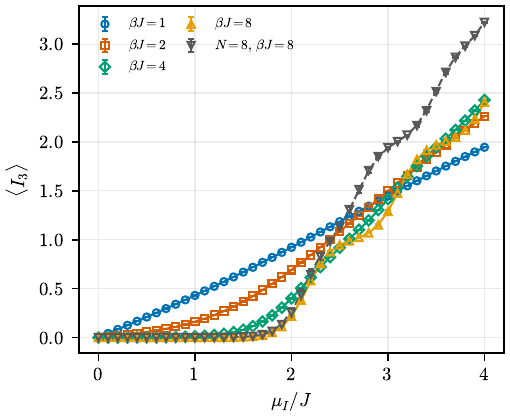}
  \caption{Two-flavor Schwinger model, $N=6$ with
  holonomy truncation $\ell_{\max}=3$. Points are from \textsc{More}, reconstructed from a single
  $\mu=0$ dataset; lines are ED.}
  \label{fig:mu}
\end{figure}

Fig.~\ref{fig:mu} shows the isospin $\langle I_3\rangle$ of the
two-flavor model at $N=6$, reconstructed across $0\le\mu_I/J\le4$ using
sector-resolved shifts. As $T\to0$, the equation of state
sharpens and Silver-Blaze behavior appears below the $T=0$ threshold
$\mu_I=2.16J$~\cite{Cohen:2003kd,Banuls:2016gid}. The $N=8$,
$\beta J=8$ result, at $\ell_{\max}=2$, provides a volume comparison.

\begin{figure}[t]
  \centering
  \includegraphics[width=\linewidth]{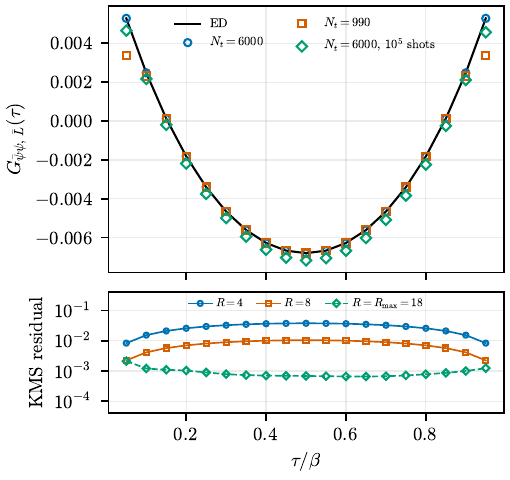}
  \caption{Euclidean condensate--electric-field correlator (top) and KMS
  residual (bottom) for $\beta J=1$, $\theta=1$, $N=4$, and $\ell_{\max}=1$. Here
  $t_{\rm cut}=60$, $R_{\max}=18$, and each shot-noise
  series uses $10^5$ shots per circuit.}
  \label{fig:kms}
\end{figure}

To test trace sampling through Eq.~\eqref{eq:kmsrel}, Fig.~\ref{fig:kms} uses the
mixed condensate--electric-field correlator:
\begin{equation}
  G_{\bar\psi\psi,\bar L}(\tau)
  =\frac{1}{Z}\tr\!\big[e^{-(\beta-\tau)H}\bar\psi\psi\,
    e^{-\tau H}\bar L\big],
  \label{eq:condensate-efield}
\end{equation}
with $\bar L=\frac{1}{N}\sum_n L_n$, at $\beta J=1$, $\theta=1$.
The reconstruction uses the double quadrature of Eq.~\eqref{eq:kms},
which needs $N_t^2$ circuits per trace state;
the figure also shows the KMS residual
$|G_{\bar\psi\psi,\bar L}(\tau)-G_{\bar L,\bar\psi\psi}(\beta-\tau)|$.
For stochastic ensembles, the
residual decreases with the number $R$ of trace states until it reaches
the $\epsilon_{\rm stat}$ floor.

The KMS residual of Fig.~\ref{fig:kms} is blind to
$\epsilon_{\rm trunc}$ and $\epsilon_{\rm disc}$, which are shared by both
orderings: an under-resolved correlator can be wrong by tens of percent
near the endpoints while satisfying Eq.~\eqref{eq:kmsrel} to machine
precision.

\begin{figure*}[t]
  \centering
  \includegraphics[width=\textwidth]{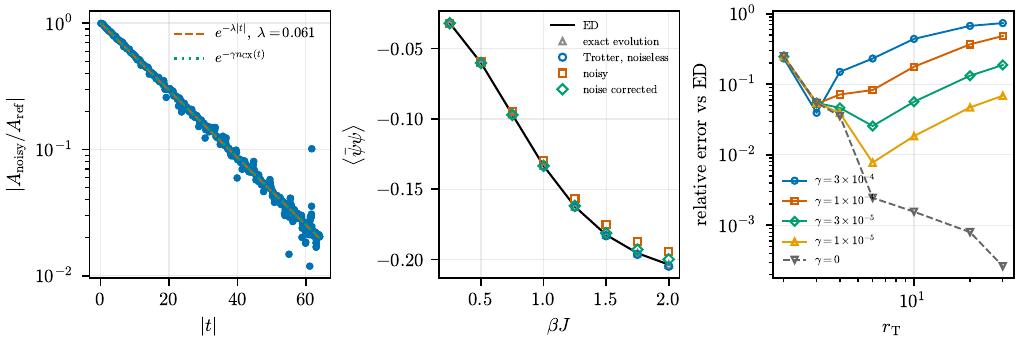}
  \caption{Circuit-level noisy simulation for $N=4$, $\ell_{\max}=1$,
  $\theta=1$, $\gamma=3\times10^{-5}$, and $r_{\rm T}=6$:
  measured damping relative to the Trotterized noiseless amplitude,
  compared with $e^{-\lambda|t|}$ and
  $e^{-\gamma n_{\rm CX}(t)}$ (left); reconstructions of $\langle\bar\psi\psi\rangle$ (center); and
  uncorrected error versus $r_{\rm T}$ for Trotterized amplitudes damped by
  $e^{-\gamma n_{\rm CX}(t)}$, averaged over $\beta J=0.5,1,2$ (right). The
  center panel uses $2\times10^4$ shots per circuit.}
  \label{fig:noisy}
\end{figure*}

Fig.~\ref{fig:noisy} treats noise as measurable broadening, through the full
circuit pipeline, comprising basis-state preparation, second-order
Trotterized Hadamard tests, and ancilla readout, executed on a
density-matrix simulator with depolarizing noise at $\gamma=10^{-5}$--$3\times10^{-4}$ per
two-qubit gate and $\gamma/20$ per one-qubit gate, with $352$ CX
per Trotter step.
Both damping parameters are fitted by least squares to
$\log|A_{\rm noisy}/A_{\rm ref}|$, with $A_{\rm noisy}$ and
$A_{\rm ref}$ the noisy and Trotterized-noiseless amplitudes measured
on the same circuits, against $|t|$ for $\lambda$ and against the
transpiled two-qubit gate count for $\gamma$.

The fit recovers $\gamma$
to $3$--$4\%$. Reweighting by
the fitted gate-count dependence reduces the error relative to the
Trotterized noiseless reconstruction by a factor of $11$--$13$ and keeps
the mean error relative to ED, before shot noise, below $1\%$ for
$\gamma\le10^{-4}$. Here $A_{\rm ref}$ is available only in simulation; on
hardware, $\gamma$ must come from calibration circuits, and the
reweighting, capped at $e^4$, inflates the variance.

Increasing $r_{\rm T}$ reduces $\epsilon_{\rm T}$ but increases
$n_{\rm CX}$, and hence the damping from hardware noise. For fixed
$\gamma$, the optimum therefore occurs at an intermediate $r_{\rm T}$.
In Fig.~\ref{fig:noisy}, the minimum error occurs at
$r_{\rm T}=3$ for $\gamma\ge10^{-4}$ and moves to $r_{\rm T}=6$ for
$\gamma\le3\times10^{-5}$.

\subsection{Fault-tolerant cost}
\label{sec:ftcost}

The total $T$-gate cost combines rotation synthesis within each circuit
with the number of circuit executions. For $N_t$ quadrature nodes, $R$
trace states, and $M$ shots for each real and imaginary amplitude,
\begin{equation}
  C_{\rm total}=2RM\sum_{t_j>0} C_T(t_j),
  \label{eq:campaign-cost}
\end{equation}
where $C_T(t_j)$ is the $T$-gate count for evolution to $t_j$; nodes with
$t_j<0$ need no circuits, since $A_n(-t)=A_n(t)^*$ for real $H$ and
$\mathcal O$. At fixed
synthesis precision, this count grows approximately with the number of
Trotter steps, $r_{\rm T}|t_j|$, and the number of non-Clifford rotations
per step.

As a fiducial example, consider the $N=10$ system of
Fig.~\ref{fig:theta} with $t_{\rm cut}=24$ and $r_{\rm T}=6$, which leaves a $3.4\%$ truncation
error at $\beta J=2$.
With the holonomy in two qubits and the product-formula terms grouped by
type, the Hadamard-test circuit, transpiled to
$\{R_z,\sqrt{X},X,\mathrm{CX}\}$, contains $9.13\times10^4$ $R_z$
rotations. Of these, $6.39\times10^4$ have arbitrary angles requiring
synthesis; the remainder are Clifford or $T$-type rotations implemented
exactly. At precision $\epsilon_{\rm syn}=10^{-10}$, each arbitrary
rotation costs about $98$ $T$ gates~\cite{Kliuchnikov:2012rs}, giving
$6.2\times10^6$ $T$ gates for an evolution to $t_{\rm cut}$. For the
actual quadrature nodes, the transpiled circuit contains
$444\,n_{\rm T}(t_j)$ arbitrary rotations. With $N_t=1884$, $R=100$,
and $M=10^4$, the campaign requires $1.9\times10^9$ executions; scaling
the shot noise of Fig.~\ref{fig:theta}, these give $\epsilon_{\rm stat}\approx0.03$ for
$\langle\bar\psi\psi\rangle$ at $\beta J=2$, before trace-sampling variance.
Summing Eq.~\eqref{eq:campaign-cost} over the nodes gives
$7.4\times10^{15}$ $T$ gates.

The electric term supplies about two thirds of the arbitrary rotations at $N=10$, through
the $O(N^2)$ charge--charge interactions left by eliminating the gauge
links; the hopping contribution grows as $O(N)$. Retaining link
registers would replace these nonlocal interactions with $O(N)$ local
terms, at the cost of one register per link. For the measured sizes
$N\le10$, the non-Clifford rotation count fits $N^{1.06}$, though its electric part
grows as $N^2$.
Extrapolating this empirical fit assumes $N_t\propto N$, whereas $N_t\sim N^{1.7}$ for $N\le10$, and fixed $R$ and
$M$ (Figs.~\ref{fig:theta} and~\ref{fig:mu} use the complete trace); both the fitted power law and these resource assumptions require
validation at larger volumes.

Five reductions are available:
\begin{enumerate}
\item Allocate shots as $M_j\propto|w_j|$, which minimizes the variance
at fixed budget. For the $N=10$ quadrature above, the estimated saving
$N_t/N_{\rm eff}$ over $\beta J=0.25$--$2$ is $4$--$33$ for temperature-specific allocations
and $4$--$8$ for a common allocation
$M_j\propto\max_\beta|w_j|$.

\item Adapt the grid. When $\Delta E$ sets $N_t$, use a uniform grid (at $N=10$, $1200$
uniform nodes converge where Gauss--Legendre needs $1500$); when the kernel does, concentrate
where it has support.

\item Replace the controlled evolution with a control-free interference
measurement~\cite{Lu:2020fine} against the empty lattice at $\ell=0$. At $N=10$ this lowers the
arbitrary rotations per Trotter step from $444$ to $215$.

\item Implement the electric energy using reversible running-charge
arithmetic rather than $O(N^2)$ pairwise Pauli
rotations. This cost is set by the charge-register
width~\cite{Kan:2025schwinger}.

\item Use catalyzed Hamming-weight phasing to synthesize the repeated
equal-angle rotations in the hopping and mass terms, allowing each disjoint
group to share a phase-gradient resource~\cite{Kan:2024hwp}.
\end{enumerate}
Together, the last two constructions reduce the asymptotic cost from
$O(N^2)$ to $O(N\log^2N)$. Composing all five reductions at $N=100$
gives an estimated saving of $70$--$140$ in the gate-count model used
here.

Block encoding reduces the cost further (Fig.~\ref{fig:lcu}): a
\textsc{prepare} over the quadrature weights, $O(N_t)$ Toffolis by
coherent alias sampling~\cite{Babbush:2018enc}, and a \textsc{select}
over $e^{-iHt_j}$. On a uniform grid, writing $t_j$ in binary factorizes
\textsc{select} into $\lceil\log_2N_t\rceil$ controlled blocks whose times
sum to the grid width rather than to
$\sum_{t_j>0}|t_j|$~\cite{Chowdhury:2016hs}, a reduction by $N_t/8$, or $150$
for the $1200$-node grid above. Coherence
costs two things. Rotations inside \textsc{select} become controlled,
and with phase-gradient synthesis their cost scales with the width of the
time register~\cite{Kan:2025schwinger}; computational-basis-diagonal
terms escape this, being time-controllable with a single rotation. And
the block encoding is subnormalized by $\lVert w\rVert_1$, so amplitude
estimation needs $O(\lVert w\rVert_1/\epsilon_{\rm stat})$ queries where
separate circuits need $O(\lVert w\rVert_1^2/\epsilon_{\rm stat}^2)$
shots~\cite{Chowdhury:2016hs} --- the quadratic improvement. Since
$\lVert w\rVert_1\propto e^{-c\tau}$ with $c=E_{\min}-\delta$, slack $\delta$ in
the spectral bound inflates $\lVert w\rVert_1$ by $e^{\delta\beta}$.
Compact
constructions such as FOQCS-LCU~\cite{DellaChiara:2025foqcs} may reduce
the \textsc{prepare}/\textsc{select} overhead further. At
$\epsilon_{\rm syn}=10^{-10}$, neither implementation is near-term; for
block-encoding estimates for this model see Ref.~\cite{Sakamoto:2023end}.

A $3{+}1$-dimensional estimate would combine evolution to
$t_{\max}=\beta K(\epsilon_{\rm trunc})$ with the reconstruction and
measurement costs above. Evolution estimates can draw on Trotter and post-Trotter
resource analyses~\cite{Kan:2021resources,Davoudi:2023algorithms,Davoudi:2026reduction,%
Rhodes:2024resources,Draper:2026block}
and explicit $SU(3)$ circuits~\cite{Balaji:2025circuits}. Encoding choices
include discrete subgroups~\cite{Alexandru:2019nsa,Gustafson:2024kym,OsorioPerez:2025sig},
field-space bases and gauge fixing~\cite{Bauer:2021gek,DAndrea:2023qnr,%
Grabowska:2024emw,Li:2026ppp}, loop-string-hadron and resource-efficient
encodings~\cite{Raychowdhury:2018osk,Kadam:2022ipf,Burbano:2024lsh,%
Kadam:2025trs,Haase:2020kaj,WebbMack:2026bkg}, pure-gauge truncations
~\cite{Ciavarella:2025truncations}, and orbifold formulations~\cite{Buser:2020cvn,%
Bergner:2024qjl,Halimeh:2024bth,Halimeh:2025framework,Bergner:2025zkj,%
Bergner:2026duh,Lamm:2026ether,Hanada:2026comments}.
The estimate must combine these ingredients with the costs of state preparation,
trace sampling and measurement. Finite $\theta$ can build on
its $3{+}1$-dimensional Hamiltonian formulation~\cite{Kan:2021theta}.
We leave the full resource estimate to future work. The resulting budget
covers a region of $(\beta,\tau,\mu)$, so comparisons should account for
this reuse.

\section{Outlook}
\label{sec:outlook}

\textsc{More} reconstructs thermal traces and correlators from real-time
amplitudes, enabling finite-$(T,\theta,\mu)$ physics without preparing
thermal states or sampling complex weights; the same primitive could serve
transport coefficients~\cite{Cohen:2021qmv,Turro:2024shear} and energy
correlators~\cite{Lee:2025energy}.
Scaling to non-Abelian theories requires digitization, projection onto
the physical trace, and repeated measurements.

Because the quantum primitive is time evolution, existing
circuits can be adapted. The hundred-qubit Schwinger-model state preparation
of~\cite{Farrell:2023fgd} and hadron propagation
of~\cite{Farrell:2024fjc} demonstrate these primitives at scale. Existing
data are reusable only if they contain
$\langle n|e^{-iHt}\mathcal{O}|n\rangle$ measured at quadrature times with
known kernel weights. Arbitrary evolution times and expectation values do not suffice.

Random superpositions within physical charge sectors offer an alternative
to computational-basis trace sampling. Applying the reconstruction to these
states could connect the method to thermal typicality~\cite{Davoudi:2022pcz}
and reduce trace-sampling variance, subject to the additional requirements of
state preparation and measurement.

Extending \textsc{More} to QCD remains an open direction. The main questions
are how to represent the non-Abelian gauge field, project onto the
gauge-invariant trace ensemble, and control the variance of that projection.
The $D_4$ thermal study~\cite{Ballini:2023qms} offers a first test.

\begin{acknowledgments}
I thank Peter Orth, Christopher Kane, Roel Van Beeumen, Eugene F.
Dumitrescu, and J\'er\^ome Gonthier for discussions that improved this
work, and for the uninterrupted stretches between them. I acknowledge the support by the Department of Energy through
the Fermilab QuantiSED program under the grant ``Toward Lattice QCD on
Quantum Computers''. This work was produced by Fermi Forward Discovery
Group, LLC under Contract No.\ 89243024CSC000002 with the U.S.
Department of Energy, Office of Science, Office of High Energy Physics.
Claude was used to assist with code development, drafting, and editorial
review. This work was performed in part at the Aspen
Center for Physics, which is supported by National Science Foundation
grant PHY-2210452.
\end{acknowledgments}

\bibliography{refs}

\end{document}